\documentclass[twocolumn,showpacs,preprintnumbers,amsmath,amssymb]{revtex4}

\usepackage{graphicx}
\usepackage{dcolumn}
\usepackage{dsfont}
\usepackage{bm}
\usepackage{float}
\newtheorem{theorem}{Theorem}
\newtheorem{lemma}{Lemma}

\begin{document}

\title{Simulation of single-qubit open quantum systems}

\author{Ryan Sweke}
 \email{rsweke@gmail.com}
\affiliation{Quantum Research Group, School of Physics and Chemistry,
 University of KwaZulu-Natal, Durban, 4001, South Africa}

\author{Ilya Sinayskiy}
 \email{sinayskiy@ukzn.ac.za}
\affiliation{Quantum Research Group, School of Chemistry and Physics, University of KwaZulu-Natal, Durban, 4001, South Africa and National Institute for Theoretical Physics (NITheP), KwaZulu-Natal, South Africa}

\author{Francesco Petruccione}
 \email{petruccione@ukzn.ac.za}
\affiliation{Quantum Research Group, School of Chemistry and Physics, University of KwaZulu-Natal, Durban, 4001, South Africa and National Institute for Theoretical Physics (NITheP), KwaZulu-Natal, South Africa}

\date{\today}

\begin{abstract}

A quantum algorithm is presented for the simulation of arbitrary Markovian dynamics of a qubit, described by a semigroup of single qubit quantum channels $\{T_t\}$ specified by a generator $\mathcal{L}$. This algorithm requires only $\mathcal{O}\big((||\mathcal{L}||_{(1\rightarrow 1)} t)^{3/2}/\epsilon^{1/2} \big)$ single qubit and CNOT gates and approximates the channel $T_t = e^{t\mathcal{L}}$ up to chosen accuracy $\epsilon$. Inspired by developments in Hamiltonian simulation, a decomposition and recombination technique is utilised which allows for the exploitation of recently developed methods for the approximation of arbitrary single-qubit channels. In particular, as a result of these methods the algorithm requires only a single ancilla qubit, the minimal possible dilation for a non-unitary single-qubit quantum channel.

\end{abstract}

\maketitle

\section{INTRODUCTION}\label{intro}

One of the primary motivations for the development of quantum computation is the possibility of efficiently simulating quantum systems \cite{SandersOne}-\cite{review2}, as suggested in Feynman's seminal paper on the topic \cite{Feynmann}. The natural first step towards this vision is the simulation of closed quantum systems, undergoing Hamiltonian generated unitary evolution, and over the past two decades consistent progress has been made in this field. Initially, Lloyd demonstrated a technique for the efficient simulation of Hamiltonians constructed as a tensor product of simpler Hamiltonians \cite{Lloyd}, and over time new methods and techniques have been introduced which have generalised the class of Hamiltonians which can be efficiently simulated while simultaneously tightening the relevant cost and error bounds \cite{Aharonov}-\cite{Berry2}.

However, equally as important is the development of methods for the simulation of open quantum systems \cite{franbook,nori}, crucial for enhancing our understanding of non-equilibrium dynamics and thermalisation in a wide range of systems, from damped-driven spin-boson models to complex many fermion-boson models \cite{review1, review2}. In particular, one would like to begin by simulating quantum channels, representing the most general quantum dynamics possible, and dynamical semigroups of quantum channels, which describe Markovian dynamics - continuous time processes resulting from interactions with a Markovian environment in the Born approximation \cite{Wolf}. A straightforward methodology for the simulation of these systems is instantly suggested by the Stinespring dilation theorem \cite{Stinespring}, in which one introduces an initially pure state environment, with size the square of the system size in the general case, such that one may simulate the open system dynamics of the system via Hamiltonian dynamics of the larger system-environment combination. Initially Lloyd \cite{Lloyd} conjectured that this approach may be improved by utilising environments initialised in a mixed state, but this conjecture was quickly falsfied by Terhal et al. \cite{Terhal}, who prove that in the worst case an environment of dimension $n^2$ is necessary for the simulation of $n$ dimensional quantum channels via the Stinespring dilation. 

An important early contribution was also made by Bacon et al. \cite{UnivMark}, who provide a method for decomposing the generators of Markovian evolution into simpler ``primitive" generators. In particular, they demonstrate that for the single qubit case universal simulation of Markovian dynamics requires only the ability to simulate a specific continuous one parameter family of generators, as well as the ability to implement the recombination methods of linear combination and unitary conjugation. The development of collision models \cite{Collision1,Collision2} for understanding quantum decoherence processes also suggests a constructive approach for the simulation of open quantum systems, and combining these insights with the results of Bacon et al. allowed for the development of collision model based methods for the simulation of single-qubit unital semigroups \cite{unital}, generalised phase-damping processes \cite{PD} and indivisible qubit channels \cite{indiv}.

More recently the notion of dissipative quantum computation and state preparation \cite{dcomp1} has been introduced, in which under the assumption of Markovian dynamics described by a Lindblad master equation, the interactions of a system with its environment are no longer considered destructive, but are instead utilised to drive a desired computational process. This formalism offers a natural setting for the simulation of open quantum systems and research in this direction has resulted in successful experimental demonstrations of the dissipative simulation of complex many-body spin models \cite{expreal1, expreal2}. In addition, dissipative quantum computation has allowed for alternative approaches to state preparation \cite{prep0}-\cite{prep7} and universal quantum computation \cite{KastThes, OQW}. Importantly however, it has recently been shown that dissipative quantum computing is no more powerful than the traditional circuit model - the so called ``Dissipative Church Turing Thesis" \cite{DissCT}. Specifically, it was shown that time evolution of an open quantum system can be efficiently simulated by a unitary quantum circuit of size scaling polynomially in the simulation time and size of the system.

Given these previous results we address in this paper the problem of constructing explicitly these efficient quantum circuits for the simulation of arbitrary Markovian processes within the traditional circuit model of quantum computation. In particular, we generalise into the super-operator regime recombination results, based on Suzuki-Lie-Trotter formulae \cite{Suzuki1, Suzuki2}, from recent Hamiltonian simulation approaches \cite{Berry}-\cite{wieb}. These results allow us to efficiently implement the recombination methods of Bacon et al. \cite{UnivMark}, such that in order to construct efficient quantum circuits for the simulation of arbitrary Markovian dynamics of a qubit it is only necessary to construct efficient circuits for the simulation of semigroups corresponding to the continuous one parameter family of generators defined by Bacon et al. \cite{UnivMark}. Furthermore, recently Wang et al. \cite{SolKitChan} have shown how to utilise convex properties of the set of single-qubit quantum channels \cite{AnalysisM2} to simulate any such channel via unitary circuits requiring only a single ancilla qubit, as opposed to the two-ancilla qubits required by straightforward implementations of the Stinespring dilation. We utilise these results for the construction of circuits for the simulation of the semigroups required by Bacon et al. \cite{UnivMark}, such that after recombination we obtain an explicit unitary circuit, with size scaling polynomially with respect to time, consisting only of CNOT gates and single qubit gates and requiring only a single ancilla qubit, for the simulation up to any desired accuracy of an arbitrary single-qubit quantum dynamical semigroup.

The structure of this paper is as follows: We begin by introducing the setting and rigorously defining the problem we wish to address. Following this we proceed in Section \ref{decomps} by presenting the method, introduced in \cite{UnivMark}, for the decomposition of an arbitrary generator of a single-qubit Markov semigroup. In Section \ref{recombs} we generalise results from \cite{papa} into the setting applicable for the problem addressed here, effectively demonstrating a method for the efficient recombination of the generators decomposed in Section \ref{decomps}. Finally, in Section \ref{sim} we exploit the methods introduced in \cite{SolKitChan} in order to provide explicit efficient unitary circuits for the semigroups corresponding to the generators resulting from the decomposition in Section \ref{decomps}. 

\section{PROBLEM AND SETTING}\label{setting}

Given a system with finite dimensional Hilbert space $\mathcal{H}_S = \mathbb{C}^d $, a quantum state of this system is described by a density matrix $\rho \in \mathcal{M}_{d}(\mathbb{C}) \cong \mathcal{B}(\mathcal{H}_{S})$, where $\rho \geq 0$, $\mathrm{tr}[\rho] = 1$ and $\mathcal{B}(\mathcal{H}_{S})$ is the algebra of bounded operators on $\mathcal{H}_{S}$. Quantum channels \cite{Wolf} provide the most general framework for describing the evolution of quantum states, and are given by completely positive, trace-preserving (CPT) maps,

\begin{equation}
T:\mathcal{B}(\mathcal{H}_{S}) \rightarrow \mathcal{B}(\mathcal{H}_{S}).
\end{equation}
Given any quantum channel $T$, there exists Kraus operators $\{K_{j} \in \mathcal{B}(\mathcal{H}_S)\}$, such that

\begin{equation}
T(\rho) = \sum_{j = 1}^r K_j \rho K_j^\dagger.
\end{equation}
In the above, $\sum_{j = 1}^r K_j^\dagger  K_j = \mathds{1} $ and $r = \mathrm{rank}(\tau) \leq d^2$ is the minimal number of Kraus operators, with $\tau \in \mathcal{B}(\mathcal{H}_S\otimes\mathcal{H}_S)$ the Jamiolkowski state,

\begin{equation}
\tau = (T\otimes\mathds{1}_S)|\Omega\rangle\langle\Omega|,
\end{equation}
where $\mathds{1}_S$ is the identity on $\mathcal{H}_S$ and $|\Omega\rangle \in \mathcal{H}_S\otimes\mathcal{H}_S$ is any maximally entangled state \cite{Wolf}. Furthermore, it is always possible to dilate the total Hilbert space in order to include an environment, such that the action of the channel on the system can be viewed as arising from the Hamiltonian generated unitary evolution of the total system and environment. Technically, it is always possible to introduce a dilation space $\mathcal{H}_{E}$ with $\mathrm{dim}(\mathcal{H}_E) = [\mathrm{dim}(\mathcal{H}_S)]^2$ such that there exists a unitary matrix $U \in \mathcal{M}_{d^3}(\mathbb{C})$ where

\begin{equation}\label{oprep}
T(\rho) = \mathrm{tr}_{E}\big[U(|e_0\rangle\langle e_0|\otimes\rho)U^{\dagger} \big]
\end{equation}
and $|e_0\rangle\langle e_0| \in \mathcal{H}_E$ is some initial state of the environment. However, in the case that $d$ is a factor of $\mathrm{rank}(\tau)$ then it is possible to construct a dilation with $\mathrm{dim}(\mathcal{H}_{E}) = r$ and $U \in \mathcal{M}_{dr}(\mathbb{C})$ - such a dilation space is called a \emph{minimal} dilation. Quantum channels as described above provide a complete picture of discrete time evolution. However, in this paper we are concerned with the simulation of Markovian continuous time evolutions, described by a continuous one parameter semigroup of quantum channels $\{T_t\}$ satisfying

\begin{equation}
T_tT_s = T_{t+s}, \qquad T_0 = \mathds{1},
\end{equation}
for $t\in\mathbb{R}_+$, where $\rho(t) = T_t\big(\rho(0)\big)$. Every continuous one parameter semigroup of quantum channels $\{T_t\}$ has a unique generator 

\begin{equation}
\mathcal{L}:\mathcal{B}(\mathcal{H}_{S}) \rightarrow \mathcal{B}(\mathcal{H}_{S})
\end{equation}
such that

\begin{equation}
T_t = e^{t\mathcal{L}} = \sum_{k = 0}^{\infty}\frac{t^k\mathcal{L}^k}{k!}
\end{equation}
and $\mathcal{L}$ satisfies the differential equation

\begin{equation}
\frac{d}{dt}\rho(t) = \mathcal{L}\big(\rho(t)\big),
\end{equation}
known as a master equation. Furthermore, a linear super-operator $\mathcal{L}:\mathcal{B}(\mathcal{H}_{S}) \rightarrow \mathcal{B}(\mathcal{H}_{S})$ is the generator of a continuous dynamical semigroup of quantum channels, if and only if it can be written in the form

\begin{equation}\label{GKS}
\mathcal{L}(\rho) = i[\rho,H] + \sum_{k,l =1}^{d^2-1} A_{l,k}\big([F_k,\rho F_l^\dagger] + [F_k\rho, F_l^\dagger] \big),
\end{equation}
where $H = H^\dagger \in \mathcal{M}_d(\mathbb{C})$ is Hermitian, $A \in \mathcal{M}_{d^2-1}(\mathbb{C}) $ is positive semidefinite and $\{F_i\}$ is a basis for the space of traceless matrices in $\mathcal{M}_d(\mathbb{C})$. Eq. \eqref{GKS} is known as the Gorini, Kossakowsi, Sudarshan and Lindblad form of the quantum Markov master equation and we refer to $A$ as the GKS matrix \cite{Wolf}. For the remainder of this paper we choose the basis $\{F_{i}\}$, without loss of generality, to be the normalized Pauli operators $\frac{1}{\sqrt{2}}\{\sigma_x,\sigma_y,\sigma_z\}$.

In order to quantify the error in approximations of quantum channels we will utilise the $(1\rightarrow 1)$-norm for super-operators, where in general the $(p\rightarrow q)$-norm of a super-operator is defined as \cite{Watrous}

\begin{equation}
||T||_{p\rightarrow q} := \sup_{||A||_p=1}||T(A)||_q.
\end{equation}
The $(p\rightarrow q)$-norm defined above is induced from the Schatten $p$-norm of an operator, defined as $||A||_p:= \big(\mathrm{tr}(|A|^p)\big)^{\frac{1}{p}}$. We use the $(1\rightarrow 1)$-norm as this is induced by the Schatten 1-norm, which corresponds up to a factor of 1/2 with the trace distance, $\mathrm{dist}(\rho,\sigma):=\sup_{0\leq A\leq 1}\mathrm{tr}\big(A(\rho-\sigma)\big)$, arising from a physical motivation of operational distinguishability of quantum states \cite{KastThes}. At this stage it is possible to succinctly state the problem which is addressed in this paper. \\

\noindent \textbf{Problem.} \textit{Given a continuous one parameter semigroup of single-qubit quantum channels $\{T_t\}$, generated by a generator $\mathcal{L}$, specified by a GKS matrix 
$A \geq 0 \in M_{3}(\mathbb{C})$ and a Hamiltonian $H = H^{\dagger} \in M_{2}(\mathbb{C})$, find a quantum circuit, acting on only the system qubit and a single ancilla qubit and using at most $\mathrm{poly}\big(||\mathcal{L}||_{(1\rightarrow 1)},t,1/\epsilon\big)$ single qubit and CNOT gates, that approximates the superoperator $T_t = e^{t\mathcal{L}}$ such that the maximum error in the final state, as quantified by the $1$-norm, is at most $\epsilon$.}\\

\noindent It is important to note that each member $T_t$ of an arbitrary semigroup of single-qubit channels $\{T_t\}$ is itself a single-qubit channel, and therefore in principle, using the methods of Wang et al. \cite{SolKitChan}, can be simulated within $1$-norm distance $\epsilon$ using $\mathcal{O}(\textrm{log}^{3.97}(1/\epsilon))$ gates from any specified single qubit set $S$ and one CNOT, acting on only the system qubit and a single ancilla. However in order to utilise this method, which may even be improved \cite{sk1,sk2} to require only $\mathcal{O}(\textrm{log}(1/\epsilon))$ such gates, it is necessary to first obtain a decomposition of the channel $T_t$ into a convex sum of quasi-extreme channels, which in order to do explicitly requires specification of the generator. Therefore in order to exploit these methods for the simulation of a semigroup generated by an \emph{arbitrary} generator, we utilise the decomposition/recombination strategy outlined in Section \ref{intro}. This strategy is inspired by approaches in Hamiltonian simulation \cite{Berry}-\cite{wieb} and as such we simultaneously adopt the notion of efficiency developed within that context. Due to our restriction to the single qubit case our notion of efficiency has no dependence on the system size, which remains a constant. However, as in \cite{SolKitChan}, we restrict ourselves to quantum circuits requiring only a single ancilla qubit, the smallest possible minimal dilation for a non-unitary single-qubit channel.

As we are restricting ourselves to single-qubit channels we begin by recalling some geometric properties of single qubit states \cite{AnalysisM2}. As $\{I,\sigma_x,\sigma_y,\sigma_z\}$ forms a basis for $\mathcal{M}_{2}(\mathbb{C})$, every density matrix $\rho$ can be written in this basis as $\rho = 1/2(\mathds{1} + \mathbf{r}\cdot\boldsymbol{\sigma})$ where $\boldsymbol{\sigma} = (\sigma_x,\sigma_y,\sigma_z)$ and $\mathbf{r} \in \mathbb{R}^3$ with $|\mathbf{r}|\leq 1$. Any single qubit quantum channel can then be represented in this basis by a unique $4\times 4$ matrix $M$, with the following structure:

\begin{equation}\label{affine1}
M = \begin{pmatrix}
1 & \mathbf{0} \\
\mathbf{m} & \tilde{M}\\
\end{pmatrix}
\end{equation}
where $\tilde{M}$ is a $3 \times 3$ matrix, $\mathbf{0}$ and $\mathbf{m}$ are row and column vectors respectively, and if we define 

\begin{equation}\label{affine3}
T(\rho) = \rho' = 1/2(\mathds{1} + \mathbf{r}'\cdot\boldsymbol{\sigma})
\end{equation} 
then $M$ defines an affine map via

\begin{equation}\label{affine2}
\mathbf{r}' = \tilde{M}\cdot\mathbf{r} + \mathbf{m}.
\end{equation}
At this stage we can proceed to develop the solution to the problem defined above, as per the strategy outlined in Section \ref{intro}.

\section{DECOMPOSITION OF ARBITRARY GENERATOR}\label{decomps}

As outlined in the description of our strategy, the first step is to provide a decomposition of an arbitrary generator $\mathcal{L}$, specified as per \eqref{GKS} by a GKS matrix $A \geq 0 \in \mathcal{M}_{3}(\mathbb{C})$ and a Hamiltonian $H = H^\dagger \in \mathcal{M}_{2}(\mathbb{C})$, into the combination of generators of simpler semigroups. This problem was initially addressed by Bacon et al. \cite{UnivMark} and we follow their strategy here. As $A\geq 0$ one can use the spectral decomposition to write,

\begin{equation}\label{decomp}
A = \sum_{k = 1}^3 \lambda_k A_{k},
\end{equation}
and therefore via linearity of $\mathcal{L}$

\begin{equation}
\mathcal{L} = \mathcal{L}_{H} + \sum_{k = 1}^3\lambda_{k}\mathcal{L}_{k},
\end{equation}
where

\begin{equation}
\mathcal{L}_{H}(\rho) = i[\rho,H]\label{def1} 
\end{equation}
and
\begin{equation}
\mathcal{L}_{k}(\rho) = \sum_{i,j =1}^{3} A_{k,(i,j)}\big([F_j,\rho F_i^\dagger] + [F_j\rho, F_i^\dagger] \big).
\end{equation}
Relabelling $\mathcal{L}_{0} := \mathcal{L}_{H}$ and defining $\lambda_{0} = 1$ we can then write

\begin{equation}\label{def2}
\mathcal{L} = \sum_{k = 0}^3 \lambda_k \mathcal{L}_k,
\end{equation}
giving us that,

\begin{equation}\label{prelim}
T_t = e^{t\mathcal{L}} = \mathrm{exp}\Big(t\sum_{k = 0}^3 \lambda_k \mathcal{L}_k\Big).
\end{equation}
Furthermore, defining $T^{(k)}_{t'} := e^{t'\mathcal{L}_k}$ we see via a straightforward implementation of the Lie-Trotter formula \cite{Suzuki1} that

\begin{align}
T_t  &= \lim_{n \rightarrow \infty}\Bigg[\prod_{k = 0}^3 e^{[t \lambda_k (\mathcal{L}_{k}/2)]/n}\prod_{k' = 3}^0 e^{[t \lambda_{k'} (\mathcal{L}_{k'}/2)]/n}\Bigg]^n\\ 
&=\lim_{n \rightarrow \infty}\Bigg[\prod_{k = 0}^3 T_{\big(\frac{t\lambda_k}{2n}\big)}^{(k)}\prod_{k' = 3}^0 T_{\big(\frac{t\lambda_{k'}}{2n}\big)}^{(k')}\Bigg]^n.
\end{align} 
Using the language of \cite{UnivMark} we say that $T_t$ can be constructed via \textit{linear combination} of the semigroups $\{T_{t'}^{(k)}\}$. In Section \ref{recombs} we present a method for the efficient recombination of linear combinations - i.e. we provide a method for the approximation of $T_t$, up to arbitrary accuracy, using only a finite (polynomial in $t$) number of implementations of channels from the constituent semigroups $\{T_{t}^{(k)}\}$. Given such a method for the efficient simulation of linear combinations, it is then clear that one can obtain an efficient algorithm for the simulation of $T_t$, provided one can efficiently simulate the constituent channels $T_{t}^{(k)}$. 

However, as per \cite{UnivMark} we can utilise basis transformations to further decompose the constituent semigroups $\{T_{t}^{(k)}\}$, and hence simplify the task of implementing channels from these semigroups, which is tackled in Section \ref{sim}. Firstly, note that for $k=1$, $\mathcal{L}_{k}$ simply generates Hamiltonian evolution, which can be simulated using a single unitary operation on a single qubit. We therefore focus on the generators of dissipative evolution, for which $k \in [2,4]$. We begin by defining \textit{unitary conjugation} of a channel $T_t$ as the procedure transforming $T_t$ according to $\mathcal{U}^{\dagger}T_t\mathcal{U}$, where $\mathcal{U}(\rho) = U\rho U^{\dagger}$ for some unitary operator $U$. Unitary conjugation preserves all Markovian semigroup properties and is clear that the effect of unitary conjugation is to apply $T_t$ in an alternative basis. In order to use unitary conjugation to further decompose the semigroups $\{T_{t}^{(k)}\}$ we utilise the following theorem, due to \cite{UnivMark}, establishing the manner in which unitary conjugation of a semigroup $\{T_t\}$ effects the GKS matrix defining the corresponding generator.

\begin{theorem}\label{universalt}
For an $N$ dimensional system, unitary conjugation of the semigroup $\{T_t\}$ by $U \in \mathrm{SU(N)}$ results in conjugation of the GKS matrix by a corresponding element in the adjoint representation of $\mathrm{SU(N)}$. 
\end{theorem}

One can then show \cite{UnivMark} that given $A_{k}$, as per \eqref{decomp}, there exists $G_k \in \mathrm{SO(3)}$, the adjoint representation of $\mathrm{SU(2)}$, such that

\begin{equation}
A_k = G_{k}A_{(\theta_k)}G_k^T,
\end{equation}
where

\begin{equation}\label{Adecomp}
A_{(\theta_{k})} = \begin{pmatrix}
\cos^2(\theta_{k}) & -i\cos(\theta_{k})\sin(\theta_{k}) & 0 \\
i\cos(\theta_{k})\sin(\theta_{k}) &\sin^2(\theta_{k}) & 0 \\
0 & 0 & 0\\
\end{pmatrix} 
\end{equation}
for $\theta_k \in [0,\pi/4]$. Therefore, as a result of Theorem \ref{universalt} there exist unitary matrices $U_k \in \mathrm{SU(2)}$ such that

\begin{equation}
T_t^{(k)}(\rho)  = U_{k}^{\dagger}\big[T^{(\theta_k)}_t \big(U_{k}\rho U_k^{\dagger} \big) \big]U_{k},
\end{equation}
where $T^{(\theta_k)}_t := e^{t\mathcal{L}_{(\theta_k)}}$ and

\begin{equation}\label{ltheta}
\mathcal{L}_{(\theta_k)}(\rho) = \sum_{i,j =1}^{3} A_{(\theta_k),(i,j)}\big([F_j,\rho F_i^\dagger] + [F_j\rho, F_i^\dagger] \big).
\end{equation}
In light of the above, we can then see that simulation of any channel from the semigroup $\{T_{t}^{(k)}\}$ requires only simulation of channels from the semigroup $\{T^{(\theta_k)}_t\}$, along with implementations of the single qubit unitary $U_{k}$.

\section{RECOMBINATION}\label{recombs}

In this section we explore the extent to which methods developed within the context of Hamiltonian simulation \cite{Berry}-\cite{wieb}, based on higher order Suzuki-Lie-Trotter integrators \cite{Suzuki1,Suzuki2}, can be generalised to construct a procedure for the simulation of $T_t$, up to arbitrary accuracy $\epsilon$, via a finite sequence of implementations of quantum channels $T^{(j)}_{t'} := e^{t'\mathcal{L}_j}$. In particular we wish to place an upper bound on the number of implementations of $T^{(j)}_{t'}$ required within this sequence. 

Given the generator $\mathcal{L} = \sum_{j=1}^m \mathcal{L}_j$ of a dynamical semigroup of quantum channels, as per \eqref{prelim} where $m=4$, we begin by assuming that 
\begin{equation}
||\mathcal{L}_1||_{1\rightarrow 1} \geq ||\mathcal{L}_2||_{1\rightarrow 1} \geq \cdots \geq ||\mathcal{L}_m||_{1\rightarrow 1}
\end{equation}
and defining the normalised component generators $\hat{\mathcal{L}}_j = \mathcal{L}_j/L_1$, where we have defined $L_j := ||\mathcal{L}_j||_{1\rightarrow 1} $ for all $j$. We then follow \cite{papa} and define the basic Lie-Trotter product formula \cite{liet,Suzuki1,Suzuki2} as,

\begin{align}
S_{2}(\hat{\mathcal{L}}_1,\ldots,\hat{\mathcal{L}}_m,\lambda) &= \prod_{j = 1}^m e^{(\frac{\lambda}{2})\hat{\mathcal{L}}_{j}}\prod_{j' = m}^1 e^{(\frac{\lambda}{2})\hat{\mathcal{L}}_{j'}}\\ \label{suz1}
&=\prod_{j = 1}^m T_{t_\lambda}^{(j)}\prod_{j' = m}^1 T_{t_\lambda}^{(j')},
\end{align}
where $t_\lambda = \lambda/(2L_1)$. Suzuki's higher order integrators  are then defined using the recursion relation

\begin{equation}\label{suz2}
S_{2k}(\lambda) = [S_{2k-2}(p_k\lambda)]^2[S_{2k-2}((1 - 4p_k)\lambda)] [S_{2k-2}(p_k\lambda)]^2,
\end{equation}
where $p_k = (4 - 4^{1/(2k-1)})^{-1}$ for $k >1$ and for notational convenience we have used $S_{2k}(\lambda)$ and $S_{2k-2}(\lambda)$ to denote $S_{2k}(\hat{\mathcal{L}}_1,\ldots,\hat{\mathcal{L}}_m,\lambda)$ and $S_{2k-2}(\hat{\mathcal{L}}_1,\ldots,\hat{\mathcal{L}}_m,\lambda)$ respectively. 

At this stage it is essential to note that for $k > 1$ we have $(1 - 4p_k) < 0$, and therefore applying the recursion rule \eqref{suz2} allows us to see that for all $k >1$ implementation of $S_{2k}(\lambda)$ requires the simulation of multiple propagators $T_{\tilde{t}}^{(j)}$ with $\tilde{t} < 0$ \cite{Suzuki1}. As such propagators are \emph{not} quantum channels (in particular they may violate complete positivity, or even positivity) \cite{franbook}, we therefore restrict ourselves here to first order ($k = 1$) integrators. This is in juxtaposition to the Hamiltonian simulation case, where for generators $\mathcal{L}_j(\cdot) = -i[H_j,\cdot]$ of purely coherent evolution, the propagators $T_{\tilde{t}}^{(j)} = e^{\tilde{t}\mathcal{L}_j}$ are quantum channels (in fact unitary conjugations) even for the case of $\tilde{t} <0$.

In light of these considerations, we therefore proceed to examine the efficiency of approximating $T_t = \mathrm{exp}(t\sum_{j = 1}^m\mathcal{L}_j)$ with sequences of quantum channels of the form $[S_2(t/r)]^x$. In particular, we note that $S_2(\lambda)$ consists of the product of $2m-1$ exponentials, and hence we can define

\begin{equation}\label{nexp}
N_{\mathrm{exp}} = (2m-1)x
\end{equation}
as the number of exponentials, and hence quantum channels, in the expression $[S_2(t/r)]^x$. The desired bound on $N_\mathrm{exp}$ is then expressed in the following theorem, a direct generalization of the work in \cite{papa} to the superoperator setting, restricted to the case of $k = 1$ in light of the above considerations.

\begin{theorem}\label{baset}
Let $1 \geq \epsilon > 0$ be such that $(9/2)L_2mt \geq \epsilon$, then for 

\begin{equation}
r \geq \frac{\sqrt{2L_2}(mt)^{3/2}}{\epsilon^{1/2}},
\end{equation}
we have that

\begin{equation}\label{bound}
\Big|\Big|\mathrm{exp}\Big( t\sum_{j = 1}^m\mathcal{L}_j\Big) - \big[S_{2}(\hat{\mathcal{L}}_1,\ldots,\hat{\mathcal{L}}_m,t/r) \big]^{rL_1}\Big|\Big|_{1\rightarrow 1}\leq \epsilon,
\end{equation}
and the number of exponentials required is bounded by

\begin{equation}\label{res}
\mathrm{N_{exp}} \leq (2m-1)  \frac{\sqrt{2L_2}L_1(mt)^{3/2}}{\epsilon^{1/2}}.
\end{equation}
\end{theorem}

In order to prove Theorem \ref{baset} we first note that the following lemma, a restriction to the case $k = 1$ of the analagous lemma in \cite{papa}, can be proven using the exact same proof as described in \cite{papa}, provided one replaces the $1$-norm with the $(1\rightarrow 1)$ norm and notes that $||T||_{1\rightarrow 1} = 1$ for any quantum channel $T$ \cite{Watrous, DissCT}, as the proof relies only on properties of the Taylor expansion of exponentials and generic properties of the norm, which hold for both Schatten norms and the induced superoperator norms \cite{Watrous}.

\begin{lemma}
For $(2/3)m|\lambda| \leq 1$ and

\begin{equation}
||\hat{\mathcal{L}}_m||_{1\rightarrow 1} \leq \cdots \leq ||\hat{\mathcal{L}}_2||_{1\rightarrow 1} \leq ||\hat{\mathcal{L}}_1||_{1\rightarrow 1} = 1,
\end{equation}
we have that

\begin{equation}
\Big|\Big|\mathrm{exp}\Big( \lambda\sum_{j = 1}^m\hat{\mathcal{L}}_j\Big) - S_{2}(\lambda)\Big|\Big|_{1\rightarrow 1} \leq 2\| \hat{\mathcal{L}}_2\|_{1\rightarrow 1}(m| \lambda |)^3,
\end{equation}
where $S_{2}(\lambda) = S_{2}(\hat{\mathcal{L}}_1,\ldots,\hat{\mathcal{L}}_m,\lambda)$.
\end{lemma}

In addition to Lemma 1, the following lemma is required:

\begin{lemma}
Given quantum channels $T$ and $V$ we have that

\begin{equation}
\big|\big| T^n - V^n \big|\big|_{1\rightarrow 1}\leq n \big|\big| T - V \big|\big|_{1\rightarrow 1}.
\end{equation}

\end{lemma}
Lemma 2 is a direct generalisation to the superoperator setting of an important result describing the accumulation of errors due to gate approximations in unitary circuits. However, in the conventional operator setting the proof relies crucially on properties of Hermitian operators and as a result an alternative proof is required within this more general setting.\\

\noindent \textit{Proof (of Lemma 2)}. It is clear that in the case that $n =1$ the lemma is true. Assume the lemma holds for $n=m$. We now show that it holds for $n = m+1$ and as a result prove the result by induction.

\begin{align}
\big|\big| T^{m+1} - V^{m+1} \big|\big|_{1\rightarrow 1}&= \big|\big|TT^m - TV^m +\nonumber\\
&\qquad\qquad TV^m - V.V^m    \big|\big|_{1\rightarrow 1}    \\
&\leq    \big|\big|T\big(T^m -V^m \big) \big|\big|_{1\rightarrow 1} + \nonumber\\
&\qquad\qquad\big|\big|\big(T -V \big)V^m \big|\big|_{1\rightarrow 1} \label{bs}\\
&\leq   \big|\big|T\big|\big|_{1\rightarrow 1} \big|\big|T^m -V^m  \big|\big|_{1\rightarrow 1} + \nonumber\\
&\qquad\qquad\big|\big|T -V  \big|\big|_{1\rightarrow 1}\big|\big|V^m\big|\big|_{1\rightarrow 1} \label{as}\\
&\leq \big|\big|T^m -V^m \big|\big|_{1\rightarrow 1} + \nonumber\\
&\qquad\qquad\big|\big|T -V \big|\big|_{1\rightarrow 1} \label{bs2}\\
&\leq (m+1)\big|\big|T -V \big|\big|_{1\rightarrow 1}
\end{align}
In the above \eqref{as} follows from \eqref{bs} via submultiplicativity of the norm, and \eqref{bs2} follows from \eqref{as} due to the fact \cite{Watrous, DissCT} that for any quantum channel $T$ we have that $||T||_{1\rightarrow 1} = 1$.  $\square$\\

\noindent Given these two lemma's it is now possible to follow \cite{papa} in order to prove Theorem 2.\\

\noindent \textit{Proof (of Theorem 2)}. First note that

\begin{equation}
\mathrm{exp}\Big( t\sum_{j = 1}^m\mathcal{L}_j\Big) = \Big[\mathrm{exp}\Big( \frac{t}{r}\sum_{j = 1}^m\hat{\mathcal{L}}_j\Big)\Big]^{rL_1},
\end{equation}
and as a result we can utilise Lemma 1 and Lemma 2 to obtain

\begin{equation}
\Big|\Big|\mathrm{exp}\Big( t\sum_{j = 1}^m\mathcal{L}_j\Big) - \big[S_{2}(\frac{t}{r})\big]^{rL_1}\Big|\Big|_{1\rightarrow 1} \leq \frac{2L_2(mt)^3}{r^2}.
\end{equation}
Therefore taking

\begin{equation}
r \geq \frac{\sqrt{2L_2}(mt)^{3/2}}{\epsilon^{1/2}},
\end{equation}
and enforcing that $(9/2)L_2mt \geq \epsilon$, ensures that both the conditions of Lemma 1 and the bound \eqref{bound} are satisfied. Finally, from eq. \eqref{nexp} one can then see that the total number of exponentials required is indeed as stated in eq. \eqref{res}. $\square$\\
\noindent

Furthermore, by definition of the $(1\rightarrow 1)$ norm we have that for any density matrix $\rho$ and any superoperators $P$ and $Q$,

\begin{equation}
||P(\rho) - Q(\rho)||_1  \leq ||P-Q||_{1\rightarrow 1}
\end{equation}
and as such the results of Theorem \ref{baset} bound the error in the output state obtained when approximating $T_t$ with $[S_{2}(t/r)]^{rL_1}$. At this point we have then established that any channel $T_t$, a member of the semigroup $\{T_t\}$ generated by $\mathcal{L} = \sum_{j=1}^m\mathcal{L}_{j}$, can be simulated up to arbitrary accuracy using $\mathcal{O}\big((L_1 t)^{3/2}/\epsilon^{1/2}\big)$ implementations of quantum channels $T_{t'}^{(j)} = e^{t'\mathcal{L}_{j}}$.

\section{SIMULATION OF CONSTITUENT SEMIGROUPS}\label{sim}

Given the results of Section \ref{decomps} and Section \ref{recombs}, all that remains is to illustrate a method for the construction of unitary circuits, consisting only of single-qubit and CNOT gates and requiring only a single ancilla qubit, for the exact implementation of quantum channels from the semigroups $\{T^{(\theta_k)}_t\}$. We proceed by following the strategy, introduced in \cite{SolKitChan}, of decomposing the channels $T^{(\theta_k)}_t$ into the convex sum of quasi-extreme channels. These quasi-extreme channels require only two Kraus operators for implementation, and hence can be simulated using a unitary circuit acting on only a single ancilla qubit. Furthermore, given a decomposition of $T^{(\theta_k)}_t$ into the convex sum of quasi-extreme channels, $T^{(\theta_k)}_t$ can be simulated using classical random sampling of these channels.

In order to obtain this convex decomposition we proceed via the following steps: Firstly, we utilise the damping basis \cite{QuMaQubits, dampingbasis} in order to find the affine map representation of $T^{(\theta_k)}_t$. From this affine map representation it is then easy to construct the Jamiolkowski state, from which it is possible to obtain the desired convex decomposition \cite{AnalysisM2}.

Using damping basis methods \cite{QuMaQubits, dampingbasis} (details can be found in Appendix A) we find, as per \eqref{affine1}-\eqref{affine2}, that the affine map representation $M$ of $T^{(\theta_k)}_t$ is given by

\begin{equation}\label{aff1}
M = \begin{pmatrix}
1 & 0 &0 & 0\\
0 & \Lambda_1 & 0 &0 \\
0 & 0 & \Lambda_2 & 0 \\
m_3 &0 & 0 & \Lambda_3\\
\end{pmatrix},
\end{equation}
where

\begin{align}
\Lambda_1 & = e^{(-2\sin^2(\theta_k)t)},\\
\Lambda_2 & = e^{(-2\cos^2(\theta_k)t)},\\
\Lambda_3 & = e^{(-2t)},\\
m_3 & = \sin(2\theta_k)(\Lambda_3 - 1). \label{aff2}
\end{align}
Given this affine representation of $T^{(\theta_k)}_t$, the Jamiolkowski state

\begin{equation}
\tau_{(\theta_k)} = (T^{(\theta_k)}_t\otimes\mathds{1}_S)|\psi_0\rangle\langle\psi_0|,
\end{equation}
with $|\psi_0\rangle = (1/\sqrt{2})(|00\rangle + |11\rangle)$, is then given by \cite{AnalysisM2}

\begin{equation}
\tau_{(\theta_k)} = \frac{1}{4}\begin{pmatrix}
a^2 &0&0&\Lambda_1 + \Lambda_2 \\
0&b^2 &\Lambda_1 - \Lambda_2&0 \\
0&\Lambda_1 - \Lambda_2 & c^2 &0 \\
\Lambda_1 + \Lambda_2 &0&0&d^2 \\
\end{pmatrix}
\end{equation}
with

\begin{align}
a &=  (1 + m_3 + \Lambda_3)^{1/2},  \\
b &=  (1 - m_3 - \Lambda_3)^{1/2},   \\
c &=  (1 + m_3 - \Lambda_3)^{1/2},  \\
d &=  (1 - m_3 + \Lambda_3)^{1/2}.   
\end{align}

In order to utilise $\tau_{(\theta_{k})}$ to obtain the desired convex decomposition of $T^{(\theta_{k})}$, we follow the procedure established in \cite{AnalysisM2}. Firstly, for any quantum channel $T$ we define $\beta(T) = 2\tau$ and note that $\beta(T)$ can always be written in the block form

\begin{equation}
\beta(T) = \begin{pmatrix}
A&C\\
C^{\dagger}&B\\
\end{pmatrix}.
\end{equation}
Furthermore, if $\hat{T}$ is the adjoint \cite{Wolf} of $T$ then 

\begin{align}
\beta(\hat{T}) &= \overline{U_{23}^{\dagger}\beta(T)U_{23}} \label{corr}\\
&= \begin{pmatrix}
A&C\\
C^{\dagger}&I-A\\
\end{pmatrix},
\end{align}
where 

\begin{equation}
U_{23} = U_{23}^{\dagger} = \begin{pmatrix}
1&0&0&0\\
0&0&1&0\\
0&1&0&0\\
0&0&0&1\\
\end{pmatrix}.
\end{equation}
Given these facts we then utilise the following three results, all due to \cite{AnalysisM2}, in order to obtain the desired convex decomposition.

\begin{theorem}\label{extremeformt}
A quantum channel $T$ is a generalised extreme point of the set of all quantum channels of the same dimension if and only if $\beta(\hat{T})$ is of the form

\begin{equation}
\beta(\hat{T}) = \begin{pmatrix}
A&\sqrt{A}U\sqrt{I-A}\\
\sqrt{I-A}U^{\dagger}\sqrt{A}& I-A\\
\end{pmatrix}.
\end{equation}
for some unitary matrix $U$.
\end{theorem}

\begin{lemma}\label{bform}
A matrix

\begin{equation}
\begin{pmatrix}
A&C\\
C^{\dagger}&B\\
\end{pmatrix}
\end{equation}
is positive semidefinite if and only if $A\geq 0 $, $B\geq 0 $ and $C = \sqrt{A}R\sqrt{B}$ for some contraction $R$. Moreover, the set of positive semidefinite matrices with fixed $A$ and $B$ is a convex set whose extreme points satisfy $C = \sqrt{A}U\sqrt{B}$ for some unitary matrix $U$.
\end{lemma}

\begin{lemma}\label{contraction}
Any contraction in $M_{2}(\mathbb{C})$ can be written as the convex combination of two unitary matrices.
\end{lemma}

In light of the above three results, our strategy for obtaining a convex decomposition of an arbitrary channel $T$ is as follows: Given $\beta(T)$ we find $\beta(\hat{T})$ using \eqref{corr}. As $T$ is completely positive this ensures that $\beta(\hat{T})\geq 0$ and as such we write $\beta(\hat{T})$ in the form described in Lemma \ref{bform}. As $R$ is a contraction we know, via Lemma \ref{contraction}, that $R$ can be decomposed into the convex combination of two unitary matrices, and as a result we obtain that 

\begin{equation}
\beta(\hat{T}) = \frac{1}{2}\beta(\hat{T}_{1}) + \frac{1}{2}\beta(\hat{T}_{2}),
\end{equation}
where due to Theorem \ref{extremeformt} we see that $T_{1}$ and $T_{2}$ are quasi-extreme channels (generalised extreme points of the set of quantum channels) providing the desired convex decomposition of $T$. Following these steps for $T^{(\theta_{k})}_t$ we find that

\begin{equation}
\beta\big(\hat{T}^{(\theta_{k})}_t\big) = \frac{1}{2}\beta\big(\hat{T}^{\theta_{k}}_{(t,1)}\big) + \frac{1}{2}\beta\big(\hat{T}^{\theta_{k}}_{(t,2)}\big),
\end{equation}

where
\begin{equation}
\beta\big(\hat{T}^{\theta_{k}}_{(t,i)}\big) = \begin{pmatrix}
A&\sqrt{A}U_i\sqrt{I-A}\\
\sqrt{I-A}U_i^{\dagger}\sqrt{A}& I-A\\
\end{pmatrix},
\end{equation}
with

\begin{equation}
U_1 = \begin{pmatrix}
0&e^{i\phi_1}\\
e^{i\phi_2}&0\\
\end{pmatrix},\qquad
U_2 = \begin{pmatrix}
0&e^{-i\phi_1}\\
e^{-i\phi_2}&0\\
\end{pmatrix},
\end{equation}

\begin{align}
\phi_1 &= \arccos\Big(\frac{\Lambda_1 + \Lambda_2}{ad}\Big),\\
\phi_2 &= \arccos\Big(\frac{\Lambda_1 - \Lambda_2}{bc}\Big),
\end{align}
and

\begin{equation}
A = \frac{1}{2}\begin{pmatrix}
a^2&0\\
0&c^2\\
\end{pmatrix}.
\end{equation}

As in \cite{SolKitChan}, in order to construct the unitary circuits implementing $T^{\theta_{k}}_{(t,i)}$ it is necessary to first find the Kraus operators $K^{i}_{1}$ and $K_{2}^{i}$, where

\begin{equation}
T^{\theta_{k}}_{(t,i)}(\rho) = \sum_{j = 1}^2 \big(K^i_j\big) \rho \big(K^i_j\big)^\dagger.
\end{equation} 
To find these Kraus operators one then uses \eqref{corr} to find the relevant Jamiolkowski state, before exploiting the standard Choi-Jamiolkowski correspondence \cite{Wolf}. Following these steps one obtains

\begin{equation}\label{Kraus1}
K_{1}^{1} = \frac{1}{\sqrt{2}}\begin{pmatrix}
0&c\\
be^{i\phi_2}&0
\end{pmatrix}\qquad
K_{2}^{1} = \frac{1}{\sqrt{2}}\begin{pmatrix}
ae^{-i\phi_1}&0\\
0&d
\end{pmatrix}
\end{equation}
and

\begin{equation}\label{Kraus2}
K_{1}^{2} = \frac{1}{\sqrt{2}}\begin{pmatrix}
0&c\\
be^{-i\phi_2}&0
\end{pmatrix}\qquad
K_{2}^{2} = \frac{1}{\sqrt{2}}\begin{pmatrix}
ae^{i\phi_1}&0\\
0&d
\end{pmatrix}.
\end{equation}
Given these Kraus operators it is then possible to find a constant size unitary circuit implementing $T^{\theta_{k}}_{(t,i)}$, consisting only of CNOT's and single qubit gates, in a variety of ways. A first method is to apply directly the results of \cite{SolKitChan} (requiring an additional two basis transformations), or alternatively one can construct from the Kraus operators unitary matrices $U^{(\theta_{k})}_{i}$, such that

\begin{equation}\label{figureref}
T^{\theta_{k}}_{(t,i)}(\rho) = \mathrm{tr}_{E}\big[\big(U^{(\theta_{k})}_{i}\big)(|0\rangle\langle 0|\otimes\rho)\big(U^{(\theta_{k})}_{i}\big)^{\dagger} \big],
\end{equation}
and proceed by obtaining a circuit decomposition of these unitary matrices. We provide an explicit demonstration of the latter strategy here. It is important to note that these unitary matrices are \emph{not} unique \cite{Wolf}, however for the purposes of this paper we choose to work with the following  form for the unitary $U^{(\theta_{k})}_{1}$,

\begin{equation}
U^{(\theta_{k})}_{1} = \begin{pmatrix}
e^{-i\phi_1}\cos(\beta) &0 &0 & -e^{-i\phi_2}\sin(\beta)\\
0&\cos(\alpha)& -\sin(\alpha) &0\\
0&\sin(\alpha)& \cos(\alpha) &0\\
e^{i\phi_2}\sin(\beta) &0 &0 & e^{i\phi_1}\cos(\beta)\\
\end{pmatrix},
\end{equation}
where we have written

\begin{align}
\cos(\beta) &= \frac{1}{\sqrt{2}}a, \qquad \sin(\beta) = \frac{1}{\sqrt{2}}b, \\
\cos(\alpha) &= \frac{1}{\sqrt{2}}d, \qquad \sin(\alpha) = \frac{1}{\sqrt{2}}c,
\end{align}
as a result of the observation that $a^2 + b^2 = 2$ and $c^2 + d^2 = 2$. Furthermore, note that $U^{(\theta_{k})}_{2}$ can be simply obtained by swapping the signs occurring within each exponential function in $U^{(\theta_{k})}_{1}$, and as such is not presented explicitly. In order to obtain an explicit circuit decomposition for $U^{(\theta_{k})}_{1}$ we note that we can write $U^{(\theta_{k})}_{1} = U^{(\theta_{k})}_{1,A}U^{(\theta_{k})}_{1,B}$, where $U^{(\theta_{k})}_{1,A}$ and $U^{(\theta_{k})}_{1,B}$ are the two-level unitary matrices

\begin{equation}
U^{(\theta_{k})}_{1,A} = 
\begin{pmatrix}
e^{-i\phi_1}\cos(\beta) &0 &0 & -e^{-i\phi_2}\sin(\beta)\\
0&1& 0 &0\\
0&0& 1 &0\\
e^{i\phi_2}\sin(\beta) &0 &0 & e^{i\phi_1}\cos(\beta)\\
\end{pmatrix}
\end{equation}
and

\begin{equation}
U^{(\theta_{k})}_{1,B} =\begin{pmatrix}
1 &0 &0 & 0\\
0&\cos(\alpha)& -\sin(\alpha) &0\\
0&\sin(\alpha)& \cos(\alpha) &0\\
0 &0 &0 & 1\\
\end{pmatrix}.
\end{equation}

\begin{figure}
\includegraphics[scale = 1]{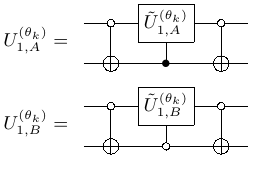} 
  \caption{Circuit decompositions for the unitary operators $U^{(\theta_{k})}_{1,A}$ and $U^{(\theta_{k})}_{1,B}$, where the unitary operator $U^{(\theta_{k})}_{1}$, implementing the quasi-extreme channel $T^{\theta_{k}}_{(t,i)}$ via \eqref{figureref}, is given by $U^{(\theta_{k})}_{1} = U^{(\theta_{k})}_{1,A}U^{(\theta_{k})}_{1,B}$. The single qubit unitary operations $\tilde{U}^{(\theta_{k})}_{1,A}$ and $\tilde{U}^{(\theta_{k})}_{1,B}$  are defined in Eqs. \eqref{p1} and \eqref{p2} respectively. }\label{C1}
\end{figure}

\begin{figure}
 \includegraphics[scale = 1]{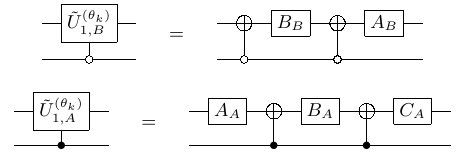} 
     \caption{Circuit decomposition for the controlled-$\tilde{U}^{(\theta_{k})}_{1,i}$ operations, required for implementation of the unitary operators $U^{(\theta_{k})}_{1,i}$, into only single qubit and controlled-NOT gates. The single qubit unitary gates are defined as $A_B = R_{y}(\alpha)$, $B_B = R_{y}(-\alpha)$, $A_A = R_{z}(\phi_1 + \phi_2)R_{y}(\beta)$, $B_A = R_{y}(-\beta)R_{z}(-\phi_1)$ and $C_A= R_{z}(-\phi_2)$, where $R_{y}(\theta)$ and $R_{z}(\theta)$ are defined in Eqs. \eqref{Ry} and \eqref{Rz} respectively.}\label{C4}
\end{figure}
Furthermore, if we define the unitary matrices,

\begin{equation}\label{p1}
\tilde{U}^{(\theta_{k})}_{1,A} = 
\begin{pmatrix}
e^{-i\phi_1}\cos(\beta) & -e^{-i\phi_2}\sin(\beta)\\
e^{i\phi_2}\sin(\beta) & e^{i\phi_1}\cos(\beta)\\
\end{pmatrix}
\end{equation}
and

\begin{equation}\label{p2}
\tilde{U}^{(\theta_{k})}_{1,B} =\begin{pmatrix}
\cos(\alpha)& -\sin(\alpha) \\
\sin(\alpha)& \cos(\alpha) \\
\end{pmatrix},
\end{equation}
then we can implement $U^{(\theta_{k})}_{1,A}$ and $U^{(\theta_{k})}_{1,B}$ using the circuits given in Figure \ref{C1}.

At this stage all that remains is to obtain circuit decompositions of the controlled-$\tilde{U}^{(\theta_{k})}_{1,i}$ gates. In order to implement the controlled-$\tilde{U}^{(\theta_{k})}_{1,B}$ gate we note the equivalence depicted in Figure \ref{C4}, where $A_B = R_{y}(\alpha)$ and $B_B = R_{y}(-\alpha)$, with $R_{y}$ the standard exponentiation of the Pauli $y$ matrix, given by

\begin{equation}\label{Ry}
R_{y}(\theta) = \begin{pmatrix}
\cos(\theta/2)&-\sin(\theta/2)\\
\sin(\theta/2)&\cos(\theta/2)\\
\end{pmatrix}.
\end{equation}
Similarly, in order to implement the controlled-$\tilde{U}^{(\theta_{k})}_{1,A}$ gate we note the equivalence depicted in Figure \ref{C4}, where $A_A = R_{z}(\phi_1 + \phi_2)R_{y}(\beta)$, $B_A = R_{y}(-\beta)R_{z}(-\phi_1)$ and $C_A= R_{z}(-\phi_2)$ with $R_{z}$ the standard exponentiation of the Pauli $z$ matrix, given by

\begin{equation}\label{Rz}
R_{z}(\theta) = \begin{pmatrix}
e^{-i\theta/2}&0\\
0&e^{i\theta/2}\\
\end{pmatrix}.
\end{equation}

\section{CONCLUSIONS AND OUTLOOK}

Combining the results of the previous three sections we obtain the following algorithm, requiring only $\mathcal{O}\big((||\mathcal{L}||_{1\rightarrow 1}t)^{3/2}/\epsilon^{1/2} \big)$ single qubit and CNOT gates, as a solution to the problem defined in Section \ref{setting}:

\begin{enumerate}
\item Given $\mathcal{L}$, obtain as per Section \ref{decomps} the spectral decomposition

\begin{equation}
\mathcal{L} = \sum_{k = 0}^4 \lambda_k \mathcal{L}_k,
\end{equation} 
as well as $G_{k}$ and $\theta_k$ specifying the decomposition

\begin{equation}
A_k = G_{k}A_{(\theta_k)}G_k^T,
\end{equation}
for all $k \in [1,3]$.
\item Choose the desired approximation accuracy $\epsilon$ as well as the simulation time $t$. Using eq. \eqref{suz1} construct $S_{2}(t/r)$ with

\begin{equation}
r =  \frac{\sqrt{2L_2}(mt)^{3/2}}{\epsilon^{1/2}},
\end{equation}

\item Implement $S_{2}(t/r)$ $L_1r$ times using

\begin{equation}
T_{t'}^{(k)}(\rho)  = U_{k}^{\dagger}\big[T^{(\theta_k)}_{t'} \big(U_{k}\rho U_k^{\dagger} \big) \big]U_{k},
\end{equation} 
where $\lambda_{k}$, $L_1$ and $r$ have been incorporated into $t'$, $U_{k}$ is obtained from $G_{k}$ as per Section \ref{decomps} and $T^{(\theta_k)}_{t'}$ is implemented via classical random sampling of the circuits derived in Section \ref{sim}.
\end{enumerate}

In light of this result two natural avenues arise for extension of this work. The first is investigation of improvements to the method presented here for the simulation of arbitrary single-qubit Markovian open quantum systems. In particular, it will be of interest to determine whether an optimality result, analogous to the ``no fast forward theorem" of Hamiltonian simulation \cite{Berry} exists in this generalised context, in which case the results of this paper would be close to optimal for the single qubit case. The second natural extension of this work is development of methods allowing for the construction of explicit algorithms for the simulation of multi-qubit and multi-qudit Markovian open systems. However, the work presented in this paper relies heavily on geometric properties of single-qubit channels and as such generalisation of this work would require investigation into the geometric and convex structure of multi-particle quantum channels, at present an open question \cite{ruskaiopen}.

 \begin{acknowledgments}
This work is based upon research supported by the South African Research Chair Initiative of the Department of Science and  Technology and National Research Foundation. Ryan Sweke acknowledges the financial support of the National Research Foundation SARChI program. The authors would also like to thank Tongyang Li and Andrew Childs for bringing to our attention problems concerning generalisation of higher-order integrators into the superoperator setting, as well as the referee for the insightful comments and suggestions, and Zehua Tian for pointing out some errors in the original manuscript.
\end{acknowledgments}

\begin{appendix}
\section{Damping basis derivation of affine map representation}

Given the generator $\mathcal{L}$ of a semigroup of quantum channels (with $H=0$) one can find the left and right eigenoperators $L_i$ and $R_i$ satisfying \cite{dampingbasis},

\begin{align}
L_i\mathcal{L} &= \lambda_{(L,i)}L_i \\
\mathcal{L}R_i &= \lambda_{(R,i)}R_i,
\end{align}
where the left action of a superoperator is defined so that

\begin{equation}
\mathrm{tr}\big[(X\mathcal{L})\rho\big] = \mathrm{tr}\big[(\mathcal{L(\rho)})X \big]
\end{equation}
for any Hermitian operator $X$ and for all density matrices $\rho$. Using this left action one finds that $\mathrm{tr}[L_iR_j] = \delta_{ij}$ and $\lambda_{(L,i)} = \lambda_{(R,i)}$. Furthermore, any density matrix $\rho(0)$ can be expressed in this basis (known as the damping basis), such that \cite{QuMaQubits}

\begin{equation}
\rho(0) = \sum_{i}\mathrm{tr}\big[L_i\rho(0)\big]R_i
\end{equation}
and

\begin{align}
\rho(t) &= e^{\mathcal{L}t}[\rho(0)] \\
&= \sum_{i}\mathrm{tr}\big[ L_i\rho(0)\big]\Lambda_i R_i
\end{align}
with $\Lambda_i = e^{\lambda_i t}$. Furthermore, the sub-matrix $\tilde{M}$ in the affine map representation of $T_t = e^{t\mathcal{L}}$ is then given by

\begin{equation}\label{subm}
\tilde{M} = \begin{pmatrix}
\Lambda_2&0&0\\
0&\Lambda_3&0\\
0&0&\Lambda_4
\end{pmatrix}.
\end{equation}
Utilising these methods for the semigroup $T^{(\theta_k)}_t$ generated by $\mathcal{L}_{(\theta_{k})}$, as per \eqref{ltheta}, we find that

\begin{align}
\lambda_2 &= -2\sin^2(\theta_k)\\
\lambda_3 &= -2\cos^2(\theta_k)\\
\lambda_4 & = -2.
\end{align}
The full affine representation, \eqref{aff1}-\eqref{aff2}, is then found using \eqref{subm} and constructing $\textbf{m}$ in  \eqref{affine1} such that \eqref{affine3} and \eqref{affine2} hold.

\end{appendix}

\end{document}